\documentstyle [12 pt] {article}
\begin {document}

\large 
\noindent {{\Large {\bf Strict spatial flatness has been  proved by the Aharonov-Bohm Effect}}}

\bigskip
\noindent {I. E. Bulyzhenkov} 
 %\affiliation %[Also at ]

 \noindent{\small  P.N. Lebedev Physical Institute, Leninsky pr. 53,  Moscow 119991, Russia\\}
 %e-mail: ibw@sci.lebedev.ru }

%\bigskip
%\noindent {\small {\bf Abstract} 
%\begin {abstract}
{\small    Abstract.{\it Would 3D space be non-Euclidean in physical reality then curved metric should destroy the Aharonov-Bohm Effect.}   }
%\end {abstract}


%\maketitle%\documentclass{article}

\bigskip

The celebrated Aharonov-Bohm phenomenon \cite {Aha} has justified not only nonlocality of overlapping electrons, but also strict spatial flatness for a distribution of electron's material density $n(x)$, which is to be a radial function. Indeed, the potential motion of any free electron in gravitational and electromagnetic fields can be formally described by a scalar charged  field    $\varphi$ = $|\varphi| exp (i\chi/\hbar c)$ in curved space-time. Therefore, one may relate,  for example \cite{Bul}, the canonical four-momentum density, $n(x) P_\mu (x) \equiv -  n(x)\nabla_\mu \chi /c = - n(x)\partial_\mu \chi /c $, of the continuous electron to the phase function gradient $\nabla_\mu \chi$ (where $\nabla_\mu$ is the covariant derivative with four space-time components, 
$\mu \rightarrow \{ 0,1,2,3\}$). The symmetrical Christoffel coefficients in Einstein's General Relativity (GR)  provide the universal covariant equality $\nabla_\mu P_\nu -
\nabla_\nu P_\mu \equiv \partial_\mu P_\nu -
\partial_\nu P_\mu \equiv -(\partial_\mu  \partial_\nu   - \partial_\nu  \partial_\mu) \chi    \equiv 0$ for charged elementary matter independently from its density function $n(x)$. However, this electron's density is to be finite in all space points where the single-valued phase $ \chi $ of matter is defined. Otherwise, $n(x) \partial_\mu \chi (x)$ would degenerate to zero under  $n(x) \equiv  0$ and the free motion of elementary matter under the phase-driven condition $n(x)\nabla_\mu P_\nu = n(x)\nabla_\nu P_\mu $ might not have any sense.
 
 For the aforesaid reason, we associate finite canonical energy-momentum densities of a distributed elementary particle (the radial electron, for example)  to all space points of the nonlocal Universe filled everywhere by gravitational and electromagnetic fields,   
\begin {equation}
 - {n_e\over c}\partial_\mu \chi_e  = n_e P_\mu \equiv   \{ n_e P_o;  n_eP_i\}   \equiv n_e g_{\mu \nu }{\left (m_o  c
{dx^\nu\over  ds} + {{e }\over c}A^\nu \right )}  $$ $$
  \equiv \left \{ \left [ {\frac {n_e m_o c {\sqrt {g_{oo}}}} {\sqrt {1 - v^2c^{-2}}}} + {  g_{o\nu}n_e eA^\nu \over c}
\right ];
 \left [ {\frac  { - n_em_o( \gamma_{ij} v^j +  {\sqrt {g_{oo}}} g_i c)} {\sqrt {1 - v^2c^{-2}} } } + { g_{i\nu}n_eeA^\nu\over c} \right ] \right \}, 
\end {equation}
where  $v^i \equiv cdx^i (g_{oo})^{- 1/2}(dx^o - g_idx^i)^{-1} \equiv dx^i /d\tau$ is the relativistic three-velocity of the local   analytical density $n_e=n_e(x)$ of the continuous charge $e = \int e n_e {\sqrt {|det \gamma_{ij}|}}dV$ and the continuous mass $m= \int m n_e {\sqrt {|det \gamma_{ij}|}}dV$, $v^2 \equiv \gamma_{ij}v^iv^j$,
 $\gamma_{ij} \equiv g_ig_jg_{oo}- g_{ij}$, $g_i \equiv - g_{oi}/g_{oo} $, $i \rightarrow \{1,2,3\}$.

Now we generalize the Sommerfeld quantization rule on the canonical four-momentum density (1) of a distributed self-coherent particle with continuous mass, $mn(x){\sqrt {|det \gamma_{ij}|}}$, and charge, $qn(x){\sqrt {|det \gamma_{ij}|}}$, densities by taking into account the GR time synchronization, 
$d\tau \equiv {\sqrt {g_{oo}}}(dx^o - g_idx^i)/c = 0$ or $dx^o = g_idx^i$, for neighboring space points of closed path-lines, 
\begin{equation}
  \pm 2\pi \hbar N = {1\over c}\oint_{d\tau = o} dx^\mu{{ \nabla_\mu \chi_s } } \equiv \oint_{d\tau = o} \left[- P_idx^i - P_o dx^o\right] 
 \equiv$$ 
{\small
$$\oint_{d\tau = o}\left[  {\frac  {m({\gamma_{ij}v^j}+{c\sqrt {  g_{oo}}}g_i)}{\sqrt {1 - v^2 c^{-2} }} } - { q(g_{io}A^o+g_{ij}A^j)\over c}-
 \left(   {\frac {mc\sqrt {g_{oo}}}  {\sqrt {1 - v^2 c^{-2} }} } + { q(g_{oo}A^o+g_{oj}A^j )\over  c}\right )g_i\right ]dx^i
$$} $$\equiv  \oint_{d\tau = o}   \left [{qA^j\over c}  + {\frac  { m{v^j }}{\sqrt {1 - v^2 c^{-2} }} }   \right ] \gamma_{ij} dx^i.
\end{equation} 
From here, the external electric potential $A^o = A^o(x)$ is not relevant to the magnetic flux quantization (2) for  instantaneous  distributions of the charged particle in electromagnetic and gravitational fields.
Indeed, the electric field $A^o(x)$ of external charges never destroyed quantization for a selected elementary particle or prevented the Aharonov-Bohm Effect in experiments with self-coherent electrons. Similarly, the gravitational field, or $g_{oo}(x)$, should also be irrelevant in practice to quantization of distributed densities of elementary matter. However, the 4D quantization rule (2) is formally assigned to gravity-dependent 3D space with the inhomogeneous metric tensor $\gamma_{ij}(x)\equiv g_ig_j g_{oo}(x) - g_{ij}(x)$. But gravity-dependent phase shifts over different lines between two material points of the same continuous electron  contradict to a single valued phase requirement for always self-coherent states of free elementary matter. Consequently, such imbalanced shifts in non-Euclidean space ought to destroy  the Aharonov-Bohm Effect. The existence of this Effect in the laboratory has confirmed strict spatial flatness of physical reality, which should be consistent with nonlocal spatial distributions of elementary sources of radial electric and gravitational fields. 

 The curved 3D metric solution was adopted for a point mass in 1916 due to then empty space paradigm. However, Einstein's metric formalism may replace the operator material density $\delta ({\bf x}- {\bf X}_e)$ of the postulated point particle with an analytical density of the continuous particle from a non-empty space paradigm.  We could propose an exact relativistic  solution $ n_e({\bf x}, t) = r_o/4\pi({\bf x}- {\bf X}_e)^2 (r_o + |{\bf x}- {\bf X}_e| )^2 $, with $r_o = Gm_o/c^2$,  for the electron's radial density in the classical (probability-free) approach to  the nonlocal elementary charge and mass around their joint center of spherical symmetry ${\bf X}_e$.  Under such a non-empty-space approach, the strongly warped 4D space-time interval for distributed radial matter gains six inherent  symmetries, $\gamma_{ij} (x) = \delta_{ij}$, which universally keep the 3D space sub-interval in nonlocal GR equations \cite{Buly, Bulyzh}. Strict spatial flatness of warped space-time manifold  reinforces the surface independent magnetic flux (2) over a closed line, the gauss electric flux over a closed surface, and gravity/inertia independent quantization of elementary matter in covariant relativistic relations \cite{Bulyz}. 
Despite the nonlocal nature of radial sources in non-empty material space, the continuous electron behaves in collisions like a material point for energy records, because half of electron's mass and charge is concentrated within the ultra small radius 
$r_o \approx 10^{-57} m$ (which is far behind the top resolution $10^{-18}m$ of space measurements).

 In 1939 Einstein first inferred the logical failure of  the Schwarzschild curved metric with singularities for physical reality from a {\it gedanken} experiment \cite {Ein}.     
Gravitation or acceleration analogs of the Aharonov-Bohm Effect were never found in SQUID experiments that indirectly confirms flatness of 3D space. The     original Aharonov-Bohm Effect has directly confirmed strict spatial flatness or $\gamma_{ij} = \delta_{ij} $ for the electron's line contour in (2).  Summing up, the 1959 Aharonov-Bohm Effect has been proved in the laboratory that has refuted non-Euclidean metric approaches to 3D distributions of nonlocal matter.

\end {document}